\documentclass[twocolumn,aps,pra,superscriptaddress,showpacs,tightenlines]{revtex4}
\usepackage{amsmath}
\usepackage{amsfonts}
\usepackage{graphicx}
\usepackage{epsfig}
\usepackage{color}
\usepackage[colorlinks,citecolor=blue]{hyperref}

\begin{document}
\title{Single-photon-induced two qubits excitation without breaking parity symmetry}
\author{Qian Bin}
\author{Xin-You L\"{u}}\email{xinyoulu@hust.edu.cn}
\author{Shang-Wu Bin}
\author{Gui-Lei Zhu}
\author{Ying Wu}\email{yingwu2@126.com}
\affiliation{School of Physics, Huazhong University of Science and Technology, Wuhan, 430074, P. R. China}
\date{\today}

\begin{abstract}
We investigate theoretically the model of two ``qubits'' system (one qubit having an auxiliary level) interacting with a single-mode resonator in the ultrastrong coupling regime.
We show that a single photon could simultaneously excite two qubits without breaking the parity symmetry of system by properly encoding the excited states of qubits. The optimal parameter regime for achieving high probability approaching one is identified in the case of ignoring the system dissipation. Moreover, using experimentally feasible parameters, we also analyze the dissipation dynamics of the system, and present the realization of two-qubit excitation induced by single-photon. This work offers an alternative approach to realize the single-photon-induced two qubits excitation, which should advance the development of single-photon quantum technologies and have potential applications in quantum information science.
\end{abstract}

\maketitle
\section{Introduction}
The interaction of light and matter has been an attractive topic in quantum optics for the last few decades~\cite{ref50}. Cavity quantum electrodynamics (QED) studies the interaction of a quantized light field with a particular boundary condition in the cavity with materials (quantum dots, natural atoms, artificial atoms, etc.), in which the quantum nature of the light field affects the dynamics of the system~\cite{ref1,ref2,ref3}. Recently, the investigations of atom-cavity  interaction have been extended to the strong coupling regime, where the coupling rate exceeds the decays of atom and cavity. Cavity QED in the strong coupling regime is very promising for the preparation and measurement of arbitrary quantum states in a fully controlled manner. It also has potential applications in the realization of quantum gates and quantum networks. Such a regime has been experimentally reached in a variety of solid systems by replacing natural atoms with artificial atoms~\cite{ref49,ref30,ref31,ref32,ref33,ref34,ref35,ref36,ref37}, giving rise to the rapidly growing of the circuit QED. In this system, superconducting qubits even can strongly interact with a single-mode resonator with a coupling rate reaching the order of $0.1$ of the field frequency, i.e., the ultrastrong coupling regime~\cite{ref42,ref43,ref44,ref45,ref55,ref60-6}. Moreover, much higher values of coupling rate have been reached in the experiments in recent years~\cite{ref60-1,ref60-2}. Recently, it is shown that three-photon resonance and single-photon-induced more atoms excitation could be realized in the ultrastrong coupling regime, which have potential applications in modern quantum technology~\cite{ref17,ref29,ref51,ref52,ref53,ref54}.

For a cavity QED system under the ultrastrong coupling regime, the method including rotating-wave approximation (RWA) is no longer able to describe exactly the dynamics of the system when higher order atom-field resonant transition is involved. The presence of counter-rotating terms in the interaction Hamiltonian can effect the collapse and revival behavior, single-scattering, and collective spontaneous emission in multiatom systems~\cite{ref12,ref13,ref14,ref15}. Physically, counter-rotating terms correspond to excitation-number-nonconserving processes involving virtual photons, which making possible multiple excitations exchange between atoms and resonators. Like ordinary quantum Rabi oscillations, this process is reversible and coherent~\cite{ref16,ref17,ref18,ref19,ref20,ref21,ref22,ref23,ref24,ref26,ref27,ref28}. In the Rabi model, the system Hamiltonian satisfies $Z^2$ symmetry due to the presence of the terms $a^\dagger\sigma^\dagger$ and $a\sigma^-$. So the state space splits into two independent subspaces or parity chains. The transitions between states induced by rotating terms and counter-rotating terms can only occur in a certain parity chain, and the transition between the two parity chains is forbidden~\cite{ref11,ref40,ref56}. In this regime, one photon can only excite odd number two-level atoms at once. Recently, it has been shown that there exists a resonant coupling between one photon and two qubits via intermediate states governed by the counter-rotating terms in the ultrastrong coupling regime~\cite{ref29}. This process needs to break parity symmetry of the atomic potentials (i.e., the $Z^2$ symmetry of Hamiltonian is broken)~\cite{ref38,ref39}, and the transition between two parity chains is realized. Then one question arises naturally. Whether two-qubit can be excited by a single photon without breaking the $Z^2$ symmetry of the system Hamiltonian.

Motivated by the above question, in this paper we introduce a circuit QED system which consists of a single-mode resonator strongly coupled to two superconducting ``qubits''. One of qubits has an auxiliary level due to its anharmonicity, which could be a phase qubit~\cite{Martinis2002,Joo2010}, capacitively shunted flux qubit~\cite{You2007,Steffen2010} or transmon qubit~\cite{Koch2007}. Under the situation that the frequency of resonator is approximately equal to the sum of the transition frequencies of two ``qubits'', we find a resonant coupling between two qubits and single-mode resonator, which allows the single-photon-induced two qubits excitation without breaking the system parity symmetry. This is because, in our proposal, the two qubit excitation state and single photon state of field are in the same parity chain by introducing the auxiliary level into a superconducting qubit with anhamonicity. We also identify the optimal parameter regime for realizing the transition between single photon state and two qubits excitation even when the system dissipation is included. Our work shows that the two qubits could also be excited by a single photon simultaneously without breaking the $Z^2$ symmetry of the system Hamiltonian, which expands its applications in quantum information science.

Our paper is organized as follows. In Sec.\,II, we introduce the model and the system dynamics in two unconnected parity chains due to the $Z^2$ symmetry of Hamiltonian.
In Sec.\,III, we demonstrate the existence of single-photon-induced two qubits excitation without breaking $Z^2$ symmetry in our proposal by diagonalizing numerically the Hamiltonian.
In Sec.\,IV, we present the single-photon-induced two qubits excitation when the system dissipation is included. In Sec.\,V, we give conclusions of our work.

\section{Model}
\begin{figure}
\includegraphics[width=8.5cm]{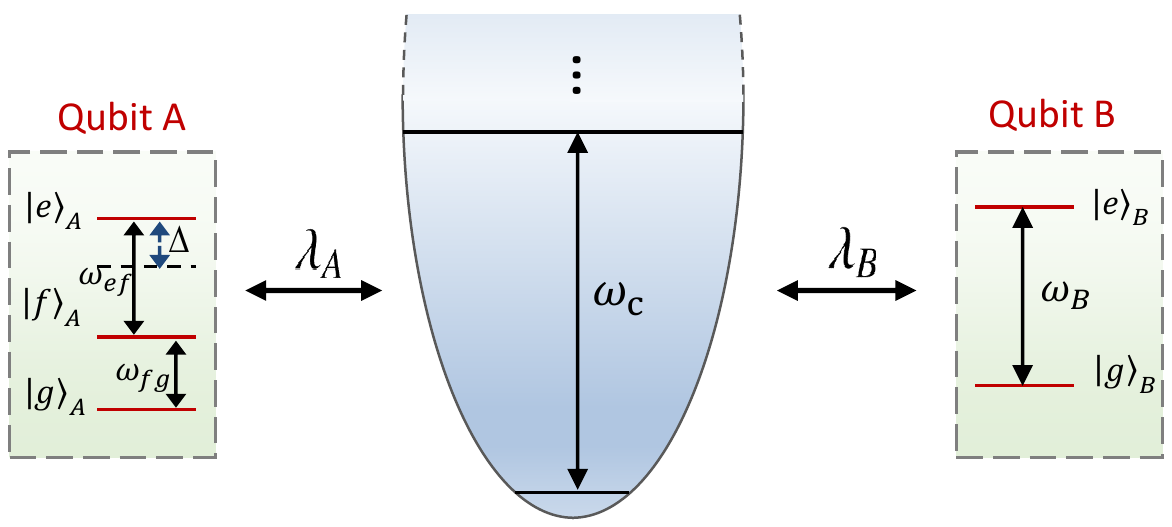}\\
\caption{Schematic of the system consists of two qubits are strongly coupled to a cavity mode. Here, $\Delta$ and $\lambda_j (j=A,B)$ are the anharmonicity of qubit A and the qubit-cavity coupling strength, respectively.
The cavity frequency is denoted by $\omega_c$, qubits frequencies are $\omega_A=\omega_{fg}+\omega_{ef}$ and $\omega_B$.}\label{fig1}
\end{figure}
As shown in Fig.\,\ref{fig1}, we consider a single-mode resonator strongly coupled to two superconducting qubits. One of the superconducting qubits has a anhamonicity $\Delta$, which leads to that three levels $|g\rangle_A$, $|f\rangle_A$, and $|e\rangle_A$ should be considered for qubit A. The Hamiltonian of this quantum system is given by $(\hbar=1)$~\cite{ref57}
\begin{equation}\label{1}
  H=H_0+H_I,
\end{equation}
where
\begin{equation}\label{2}
H_0=\omega_c a^\dag a+\omega_A|e\rangle_A\langle e|+\omega_{fg}|f\rangle_A\langle f|+\omega_B|e\rangle_B\langle e|
\end{equation}
is the free Hamiltonian of cavity mode and the two qubits. $H_I$ describes the interaction between cavity mode and two qubits with~\cite{ref59}
\begin{align}\label{3}
H_I=&\lambda_A(a+a^\dag)(\sqrt{2}|e\rangle_A\langle f|+|f\rangle_A\langle g|+h.c.)\nonumber\\
&+\lambda_B(a+a^\dag)(|e\rangle_B\langle g|+h.c.).
\end{align}
Here, $a$ and $a^\dag$ are, respectively, the annihilation and creation operator for the cavity mode with frequency $\omega_c$, $\omega_A$ and $\omega_B$ are qubits frequencies. $\omega_{fg}$$(\omega_{ef})$ is the transition frequency between the states $|f\rangle_A$ $(|e\rangle_A)$ and $|g\rangle_A$ $(|f\rangle_A)$ of the qubit A. The transition between the levels $|g\rangle_A$ and $|e\rangle_A$ is forbidden due to the fact that they have the same parities in the superconducting qubit when the reduced magnetic flux is set as $0.5$~\cite{ref38}. We consider $\omega_A=\omega_{fg}+\omega_{ef}$ and $\omega_{ef}=\omega_{fg}+\Delta$.

\begin{figure}
\includegraphics[width=8.5cm]{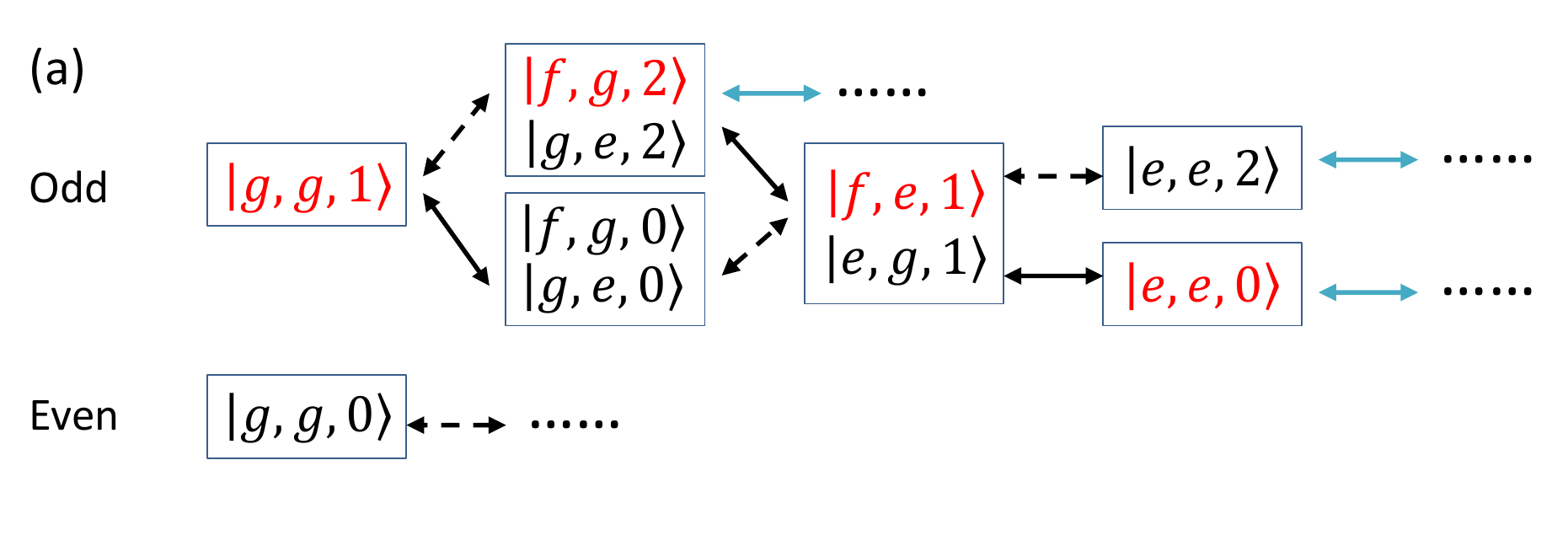}
\includegraphics[width=8cm]{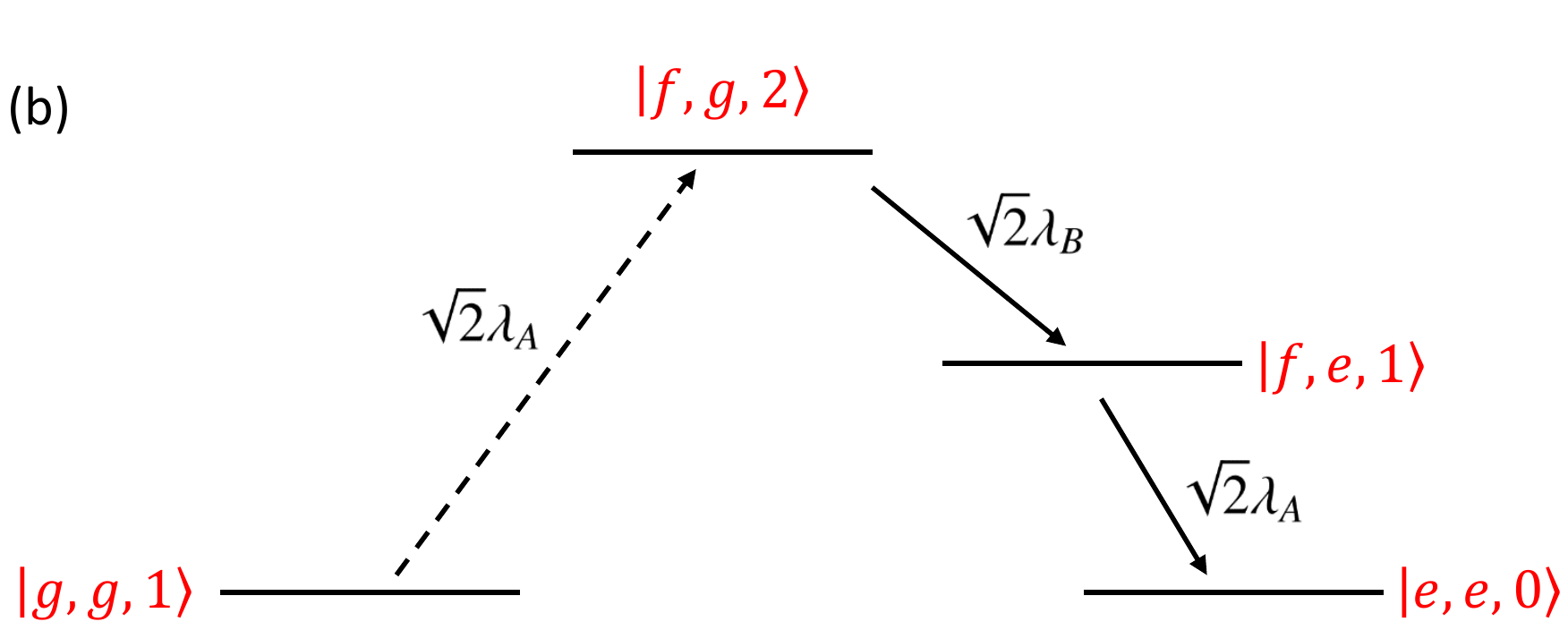}\\
\caption{ (a) Two parity chains describing the dynamical evolution of system decided by the $Z^2$ symmetry. The transition between two states in the same chain may be connected via either rotating or counterrotating terms of Hamiltonian $H$. The first, second and third indexes in state $|A,B,C\rangle$ denote qubits A, B, and resonator, respectively. (b) One of the processes of the effective coupling between the states $|g,g,1\rangle$ and $|e,e,0\rangle$ depicted by the red text in (a). Here, $|f,g,2\rangle$ and $|f,e,1\rangle$ are intermediate states, $\sqrt{2}\lambda_A$ and $\sqrt{2}\lambda_B$ are transition matrix elements, the excitation-number-nonconserving processes are represented by arrowed black dashed line.}\label{fig2}
\end{figure}
The system Hamiltonian $H$ has parity (or $Z^2$) symmetry with a well defined parity operator
\begin{equation}\label{4}
\Pi=e^{i\pi N}=e^{i\pi [a^\dag a+2|e\rangle_A\langle e|+|f\rangle_A\langle f|+|e\rangle_B\langle e|]},
\end{equation}
$\Pi |p\rangle=p |p\rangle$ $(p=\pm1)$, which measures an odd-even parity of the system dynamics, and the parity operator commutes with the Hamiltonian $H$. The system dynamics inside the Hilbert space is split into two unconnected parity chains, as displayed in Fig.\,\ref{fig2}(a)~\cite{ref11,ref40,ref56}. States within each parity chain may be connected via either rotating or counter-rotating terms. Interestingly, in our model, the single photon state $|g,g,1\rangle$ and the two qubit excitation state $|e,e,0\rangle$ are in the same parity chain, i.e., odd chain. Then the rotating or counter-rotating terms of the system Hamiltonian enable many paths for realizing the single-photon-induced two-qubit excitation, i.e., $|g,g,1\rangle\rightarrow|e,e,0\rangle$. Each path includes several virtual transitions via out of resonance intermediate states. Various coupling rates are obtained through different transition paths. Therefore all of paths should be considered in the calculation of the effective coupling rate between the states $|g,g,1\rangle$ and $|e,e,0\rangle$~\cite{ref29}. There are six paths in the lower-order transition processes, one of the transition paths is displayed in Fig.\,\ref{fig2}(b). In this transition process, the counter-rotating term $a^\dag|f\rangle_A\langle g|$ induces the transition $|g,g,1\rangle\rightarrow|f,g,2\rangle$, while the rotating terms $a|e\rangle_B\langle g|$ and $a|e\rangle_A\langle f|$, induce $|f,g,2\rangle\rightarrow|f,e,1\rangle$ and $|f,e,1\rangle\rightarrow|e,e,0\rangle$, respectively. Here, higher-order processes can also be considered, which depending on the coupling strength $\lambda$. Since there are too many intermediate states in the higher-order process, when it has the same coupling strength as the low-order process, the effective coupling rate between the states $|g,g,1\rangle$ and $|e,e,0\rangle$ is weaker than that of the latter. The auxiliary energy level $|f\rangle_A$ is considered as an excess energy level in this process. Furthermore, we find that, if both qubits have auxiliary levels, the transition between the states $|g,g,1>$ and $|e,e,0>$ can be seen in the situation that the parity symmetry of atomic potential is broken~\cite{ref29}. However, in our proposal, this transition will not happen due to both of them are in two unconnected parity chains.

\section{single-photon-induced two qubits excitation without dissipation}
\begin{figure}
  \centering
  \includegraphics[width=8.5cm]{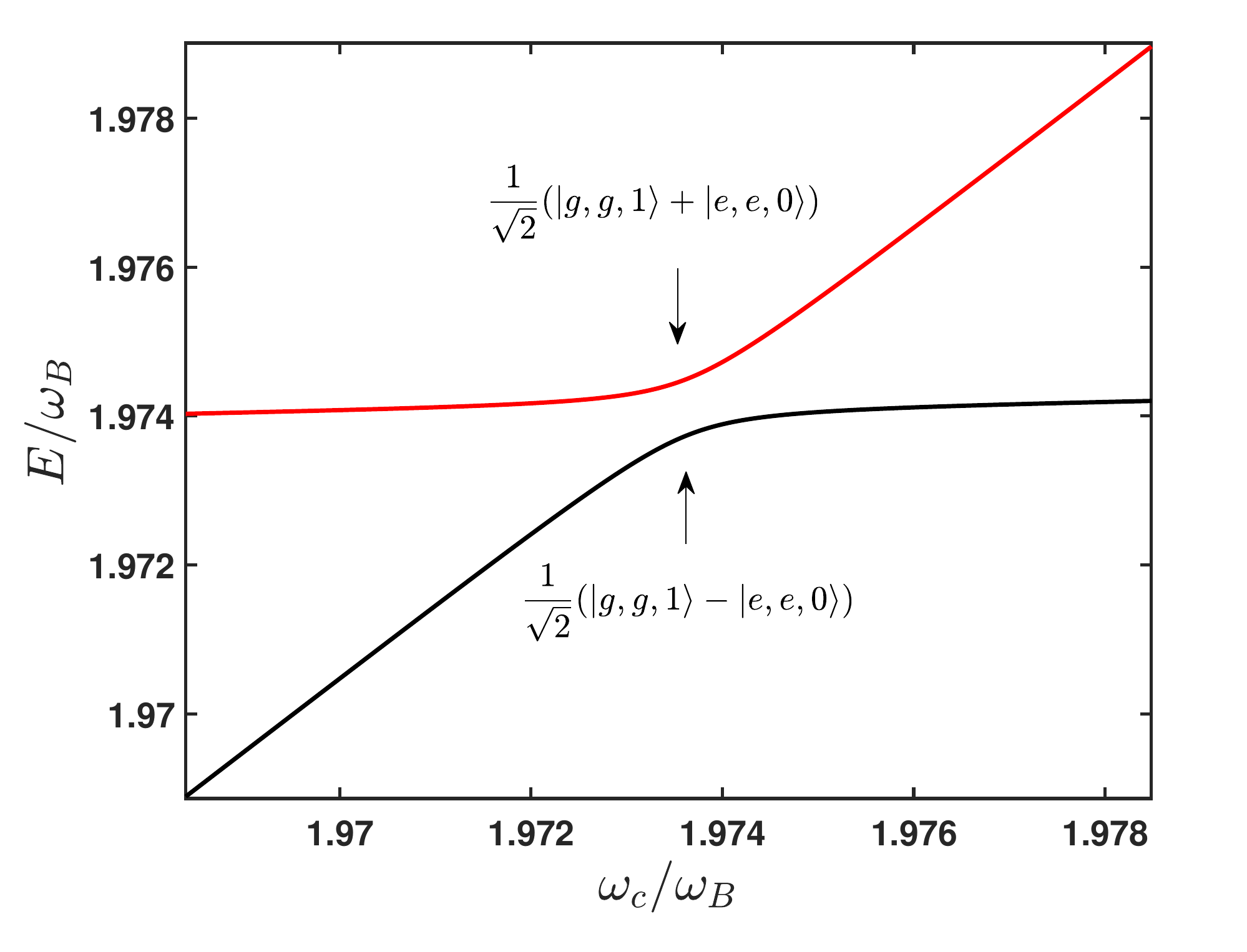}\\
  \caption{The energy levels $E_6/\omega_B$ and $E_5/\omega_B$ versus the cavity frequency $\omega_c/\omega_B$. Here, we consider a normalized coupling rate $\lambda/\omega_B = 0.1$. we used $\omega_A=\omega_B$, $\omega_{fg}=(\omega_B-\Delta)/2$ and $\Delta=0.4\omega_B$. The black arrows indicate an avoided-level crossing.}\label{fig3}
  \end{figure}

We first consider that our system is designed to operate in the regime where the qubit-cavity detuning is large compared with the qubit-cavity coupling strength. Therefore, we have $\omega_c-\omega_{B}\gg\lambda_j$ and $\omega_c-\omega_{A}\gg\lambda_j$ $(j=A,B)$. We are interested in the situation where the sum of the transition frequencies of two qubits is approximately equal to the frequency of the cavity mode, i.e., $\omega_A+\omega_B\approx\omega_c$. In this parameter region, assuming $\lambda_A=\lambda_B=\lambda=0.1\omega_B$, we can obtain the energy spectrum of the system by diagonalizing numerically the Hamiltonian in Eq.\,(1), $H|\varphi_n\rangle=E_n|\varphi_n\rangle$. We indicate a part of the dependence of energy spectrum on the single-mode resonator frequency $\omega_c$ in Fig.\,\ref{fig3}. Here, we focus on the 5th and 6th eigenstates of Hamiltonian $H$. It is shown that the two levels display an avoided crossing around $\omega_c/\omega_B=2$ with the value of the energy splitting $2\Omega_{\rm eff}$ about $7.58\times10^{-4}\omega_B$. The result is consistent with the effective coupling rate $6.83\times10^{-4}\omega_B$, obtained by the standard third-order perturbation theory, i.e., $\Omega_{\rm eff}=\frac {16\sqrt{2}}{3} \frac {\Delta\lambda^3}{\omega_B(9\omega_B^2-\Delta^2)}$ ~\cite{ref29}. This avoided crossing illustrates the presence of the resonant transition between the states $|g,g,1\rangle$ and $|e,e,0\rangle$, which demonstrates that single-photon can induce two qubits excitation in our model. Here, the system Hamiltonian satisfies $Z^2$ symmetry.

\begin{figure*}
  \centering
  \includegraphics[width=16cm]{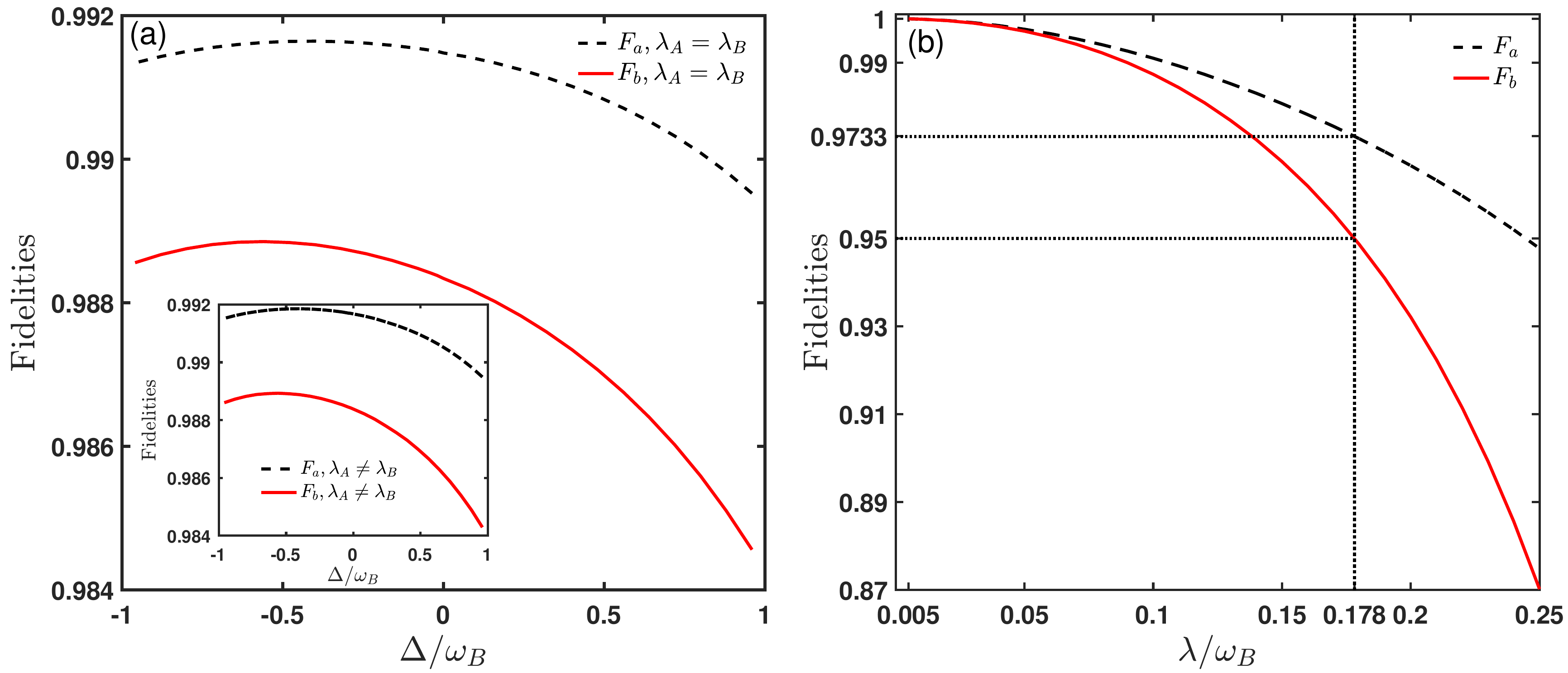}\\
  \caption{(a) The fidelities $F_a$ and $F_b$ for $\omega_c$ at the avoided-crossing point versus  $\Delta/\omega_B$ when $\lambda_A=\lambda_B=0.1\omega_B$. The inset correspond to the fidelities for $\lambda_A=0.105\omega_B$ and $\lambda_B=0.095\omega_B$. (b) The fidelities $F_a$ and $F_b$ versus $\lambda/\omega_B$ for $\omega_c$ at the avoided-crossing point.
  The system parameters used here are: (a) $\omega_A=\omega_B$, $\omega_{fg}=(\omega_B-\Delta)/2$ and $\Delta=0.4\omega_B$; (b) $\omega_A=\omega_B$, $\omega_{fg}=(\omega_B-\Delta)/2$ and $\lambda=0.1\omega_B$.}\label{fig4}
\end{figure*}
Furthermore, we find that the change of energy level $|\varphi_5\rangle$ by the inclined line part and the flat line part of it in Fig.\,\ref{fig3}, which reflects that $|\varphi_5\rangle$ changes from $|g,g,1\rangle$ to $|e,e,0\rangle$ as $\omega_c/\omega_B$ increases. For energy level $|\varphi_6\rangle$, the change is reversed. The energy splitting clearly shows the hybridizations of the states $|g,g,1\rangle$ and $|e,e,0\rangle$. Note that the hybrid states can be given approximately by $|a\rangle=(|g,g,1\rangle+|e,e,0\rangle)/\sqrt2$ and $|b\rangle=(|g,g,1\rangle-|e,e,0\rangle)/\sqrt{2}$. The fidelities $F_a=|\langle a|\varphi_n\rangle|$ and $F_b=|\langle b|\varphi_{m}\rangle|$ can get by numerically calculating the Hamiltonian $H$. Here, $|\varphi_n\rangle$ and $|\varphi_m\rangle$ are two adjacent eigenstates of Hamiltonian that maximize $F_a$ and $F_b$. In Fig.\,\ref{fig4}(a), we show that the dependence of the fidelities $F_a$ and $F_b$ on $\Delta/\omega_B$ when $\lambda_A=\lambda_B $ (the main part) and $\lambda_A\neq\lambda_B$ (the insert). Comparing the main part and insert of Fig. 4(a), it is shown that our results are robust to the deviation in the couplings between the qubits and the resonator. Therefore, we could always consider $\lambda_A=\lambda_B=\lambda$ for later discussion. Moreover, we find that the fidelities $F_a$ and $F_b$ both are greater than 0.984 within the given parameter range. Fig.\,\ref{fig4}(b) shows that both $F_a$ and $F_b$ are greater than 0.95 for $\lambda/\omega_B<0.178$. The bigger $\lambda/\omega_B$ is, the lower the fidelities are. This is because when the interaction strength is too high, the eigenstates of the system Hamiltonian will be seriously dressed due to the nonlinearity of the qubits so that they can not be approximately given by bare states~\cite{ref17}. For a weaker value of $\lambda/\omega_B$, a high fidelity is obtained. But it also makes the effective coupling strength $2\Omega_{\rm eff}/\omega_B$ smaller, which corresponds to a small transition speed between $|g,g,1\rangle$ and $|e,e,0\rangle$. So it is not beneficial to have a very small value of $\lambda/\omega_B$. In the limit of $\lambda/\omega_B=0$, there is no coupling between the qubits and resonator. Then the transition between the initial state and target state will disappear. Based on the definition of fidelities, the corresponding fidelities are $F_a=F_b=1/\sqrt{2}$ in this case.

\section{single-photon-induced two qubits excitation with dissipation}
We consider the coupling of the system to the environment with the interaction Hamiltonian
\begin{equation}\label{5}
H_{sb}=\sum\limits_{\mu,\nu}\alpha_{\mu,\nu}(s_\mu+s^\dag_\mu)(b_{\mu\nu}+b^\dag_{\mu\nu}),
\end{equation}
where $\alpha_{\mu,\nu}$ is the system-bath coupling strength, $\mu$ marks the cavity mode or the transitions of the qubits. $s_\mu$ is the photon operator or the lowering operator in the system, and $b_{\mu\nu}$ $(b^\dag_{\mu\nu})$ is the annihilation (creation) operator for the bath mode $\mu\nu$ with frequency $\omega_\nu$~\cite{ref23}. In order to describe properly the dynamics of the system, we calculate the influence of decay of cavity field and qubits on the time evolution of the mean photon number and the two-qubit occupation probability in their excited states. We adopt the master equation approach to calculate the dynamics of the system. By applying the standard Markov approximation, we derive the master equation for tracing out the environment degrees of the freedom,
\begin{align}\label{6}
\frac {d\rho}{dt}=i[\rho,H]+\kappa\mathcal{L}[a]\rho+\gamma_{ef}^A\mathcal{L}[|f\rangle_A\langle e|]\rho\\
+\gamma_{fg}^A\mathcal{L}[|g\rangle_A\langle f|]\rho+\gamma_{eg}^B\mathcal{L}[|g\rangle_B\langle e|]\rho,
\end{align}
with the superoperator $\mathcal{L}$ is expressed as
\begin{equation}\label{7}
\mathcal{L}[\emph{O}]=\frac {1}{2}(2\emph{O}\rho\emph{O}^\dag-\rho\emph{O}^\dag\emph{O}-\emph{O}^\dag\emph{O}\rho).
\end{equation}
Here, we write the system-bath interaction Hamiltonian $H_{sb}$ in the basis formed by the eigenstates of $H$, $E_m>E_n$ for $m>n$. We consider the temperature of environment to be zero $(T=0)$.
For simplicity, we assume $\alpha_{\mu,\nu}$ and the spectral density of bath at the transition frequency to be constant. $\kappa$ is the decay rate of the single-mode cavity, $\gamma_{l,k}^A$ $(l, k=e, f, g)$ is the relaxation rate the energy levels from $|l\rangle_A$ to$|k\rangle_A$, and $\gamma_{eg}^B$ is the relaxation rate the energy levels from $|e\rangle_B$ to$|g\rangle_B$.
\begin{figure}
  \centering
  \includegraphics[width=8cm]{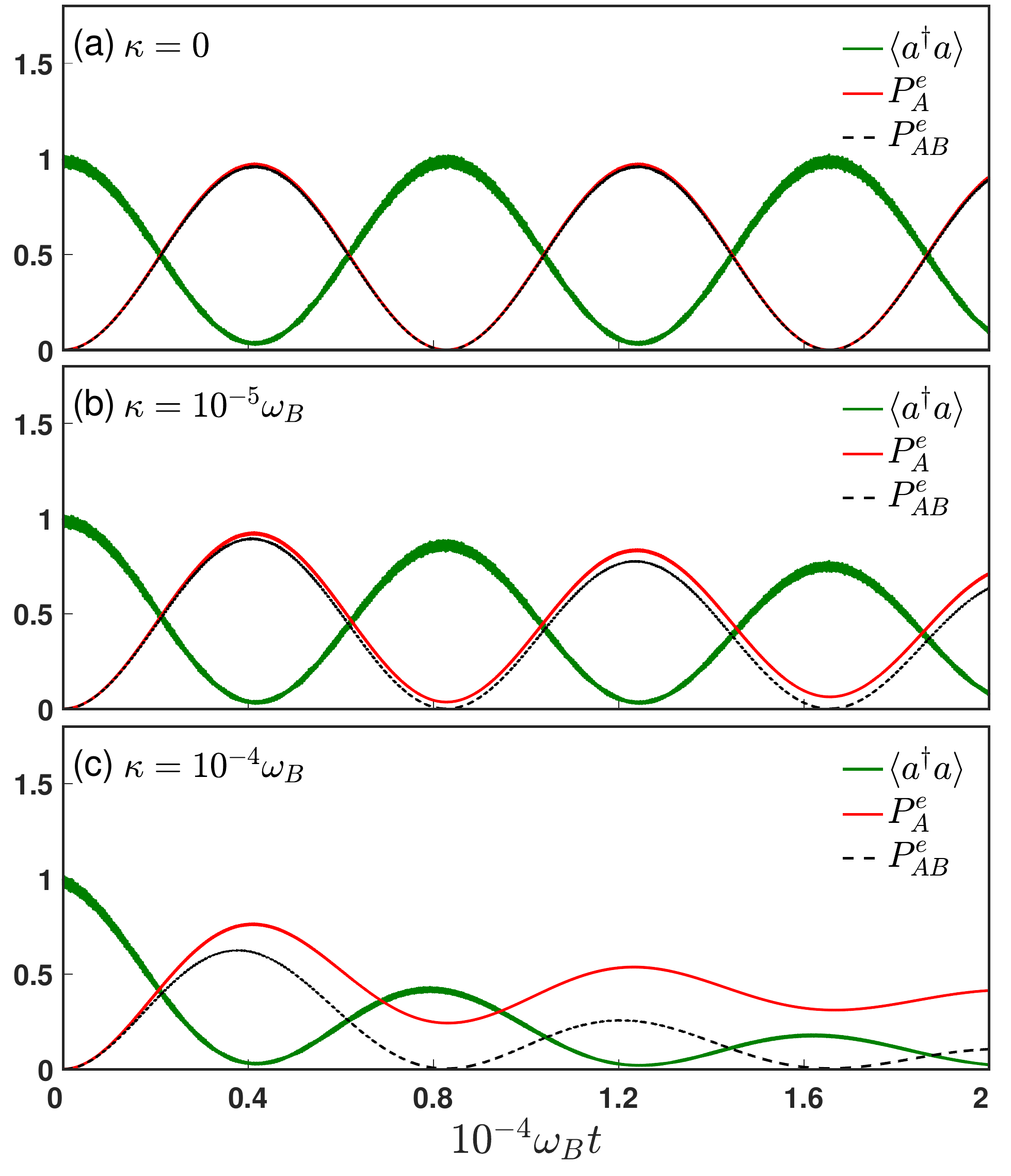}\\
  \caption{Time evolution of the expectation value of the photon number $\langle a^{\dagger} a\rangle$ (green solid curve), the occupation probability of qubit A in the excited state $|e\rangle_A$ (red solid curve), and the two-qubit occupation probability in state $|e,e\rangle_{AB}$ (black dashed curve). The system parameters used here are $\lambda/\omega_B = 0.1$, $\omega_A=\omega_B$, $\omega_{fg}=(\omega_B-\Delta)/2$, $\Delta=0.4\omega_B$, $\gamma_{fg}^A=\gamma_{ef}^A/\sqrt{2}=\gamma_{eg}^B=\kappa$; And $\kappa=0$ for panel (a), $\kappa=10^{-5}\omega_B$ for panel (b) and $\kappa=10^{-4}\omega_B$ for panel (c).}\label{fig5}
\end{figure}

In the ultrastrong coupling regime, quantum optical normal order correlation functions fail to describe photon detection experiments properly. It has been shown that, for a single-mode resonator, the output photon rate can be detected by a photoabsorber. It is proportional to $\langle X^-X^+\rangle$, i.e., $\langle X^-X^+\rangle\propto\langle a^\dag a\rangle$, where $X^+$ and $X^-$ are the positive frequency and negative frequency component of the quadrature operator $X=a+a^\dag$, respectively~\cite{ref23,ref58}. $X^+$ can be expressed as
\begin{equation}\label{9}
X^+=\sum\limits_{E_n,E_m>E_n}X_{nm}|\varphi_n\rangle\langle \varphi_m|,
\end{equation}
where $X_{nm}=\langle \varphi_n|(a^\dag+a)|\varphi_m\rangle$ and $X^-=(X^+)^\dag$. Considering the RWA or in the large-detuning limit, $X^-$ and $X^+$ coincide with $a^\dag$ and $a$, respectively.

The output photon flux coming out of the cavity can be expressed as
\begin{equation}\label{10}
\Phi=\kappa \langle X^-X^+\rangle=\kappa {\rm Tr}[\rho(t)X^-X^+].
\end{equation}

Similarly, in the ultrastrong coupling regime, the qubits emission rate can be detected by coupling the qubit to an additional microwave antenna. The result is proportional to $\langle C^-C^+\rangle$, where $C^+$ and $C^-$ are the positive frequency and negative frequency component of the qubit operator, respectively. $C^+_i$ can be expressed as
\begin{gather}\label{8}
  C^+_{i}=\sum\limits_{E_n,E_m>E_n}C_{nm}^i|\varphi_n\rangle\langle \varphi_m|,\\
  C_{nm}^1=\langle \varphi_n|(|e\rangle_A\langle f|+|f\rangle_A\langle e|)|\varphi_m\rangle, \\
  C_{nm}^2=\langle \varphi_n|(|f\rangle_A\langle g|+|g\rangle_A\langle f|)|\varphi_m\rangle,\\
  C_{nm}^3=\langle \varphi_n|(|e\rangle_B\langle g|+|g\rangle_B\langle e|)|\varphi_m\rangle,
\end{gather}
where $C^-_i=(C^+_i)^\dag$. Considering the RWA or in the large-detuning limit, $C^-_1$ ,$C^+_1$,  $C^-_2$, $C^+_2$ ,$C^-_3$ and $C^+_3$ coincide with $|e\rangle_A\langle f|$, $|f\rangle_A\langle e| $, $|f\rangle_A\langle g|$, $|g\rangle_A\langle f|$, $|e\rangle_B\langle g|$ and $|g\rangle_B\langle e|$, respectively.
The operator $X$ and $C$ both involve only the transitions from higher eigenstates $|\varphi_m\rangle$ to lower eigenstates $|\varphi_n\rangle$~\cite{ref17}.
The master equation in Eq.\,(6) can be written as
\begin{equation}\label{6}
\frac {d\rho}{dt}=i[\rho,H]+\sum\limits_{\mu}\sum\limits_{E_n,E_m>E_n}\Gamma_{\mu}^{nm}\mathcal{L}[|\varphi_n\rangle\langle \varphi_m|]\rho,
\end{equation}
with the relaxation coefficients can be given by
\begin{gather}\label{8}
  \Gamma_{1}^{nm}=\kappa|\langle \varphi_n|X|\varphi_m\rangle|^2, \\
  \Gamma_{2}^{nm}=\gamma_{ef}^A|\langle \varphi_n|(|e\rangle_A\langle f|+|f\rangle_A\langle e|)|\varphi_m\rangle|^2, \\
  \Gamma_{3}^{nm}=\gamma_{fg}^A|\langle \varphi_n|(|f\rangle_A\langle g|+|g\rangle_A\langle f|)|\varphi_m\rangle|^2,\\
  \Gamma_{4}^{nm}=\gamma_{eg}^B|\langle \varphi_n|(|e\rangle_B\langle g|+|g\rangle_B\langle e|)|\varphi_m\rangle|^2.
\end{gather}

\begin{figure}
  \centering
  \includegraphics[width=8cm]{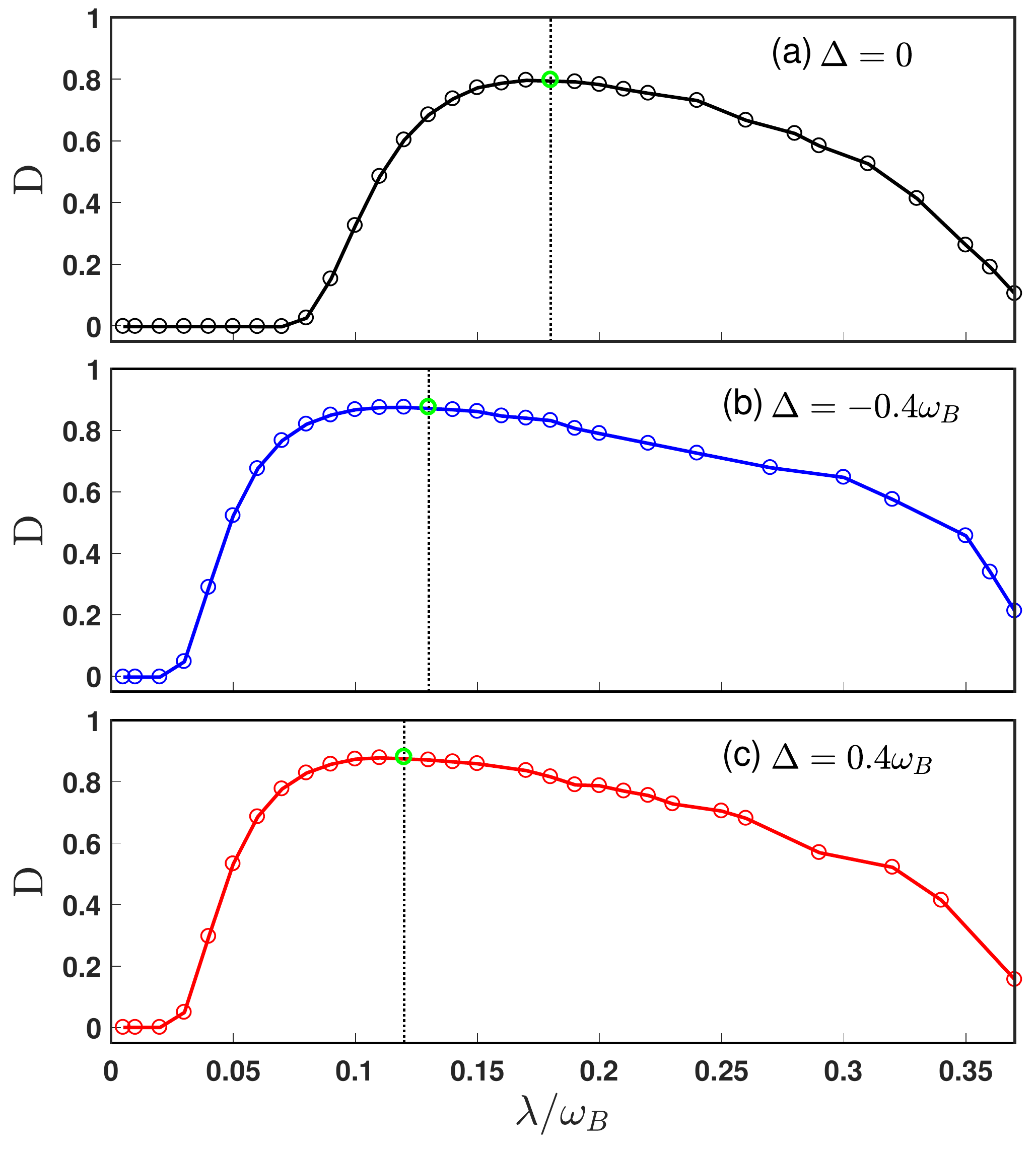}\\
  \caption{The maximum difference $D$ between the two-qubit occupation probability in their excited states $P_{AB}^e$ and the mean cavity photon number $\langle a^\dag a\rangle$ in a period of time versus $\lambda/\omega_B$. The maximum values of the curves are marked in green circles. The system parameters are the same as Figure.\,\ref{fig5}(b) except for: (a) $\Delta=0$; (b) $\Delta=-0.4\omega_B$; (c) $\Delta=0.4\omega_B$;}\label{fig6}
\end{figure}

We numerically calculated the dynamics of the system in Fig.\,\ref{fig5}. It is shown that the time evolution of the cavity mean photon number $\langle a^\dag a\rangle$, the occupation probability of qubit A in the excited state $P_A^e$, and the two-qubit occupation probability in their excited states $P_{AB}^e$. We find that the excitation exchange between the single-mode resonator and two qubits is reversible. Fig.\,\ref{fig5}(a) displays that the qubits reach fully the state $|e,e\rangle_{AB}$ with a probability approaching to one without including the cavity field damping and qubits decay. In particular, we observe that the time evolution curves of $P_A^e$ and $P_{AB}^e$ almost coincide during this process. The almost perfect two-qubit excitation demonstrates the presence of single photon induced two-qubit excitation in our model. Moreover, we also see that $\langle a^\dag a\rangle$ hardly ever goes to zero when $p_{AB}^e$ has maximal value. The reason for this behavior is that two-qubit excited state acquires a dipole transition matrix element to emit photons. So this transition process induces its coupling with single-photon state~\cite{ref29}.

The influence of cavity field damping and qubits decay on the system dynamics has been shown in Figs.\,\ref{fig5}(b) and \,\ref{fig5}(c). In contrast to Figs.\,\ref{fig5}(b) and \,\ref{fig5}(c), we find that the two-qubit excitation loss increases as the rate of cavity field damping and qubit decay increase. It can be predicted that, if the rates continue to increase, photon and qubits will not exchange excitation obviously. The reason is that the system dissipation so fast that the interaction between the cavity field and qubits is just beginning, however, the regime has collapsed heavily.

\begin{figure}
  \centering
  \includegraphics[width=8cm]{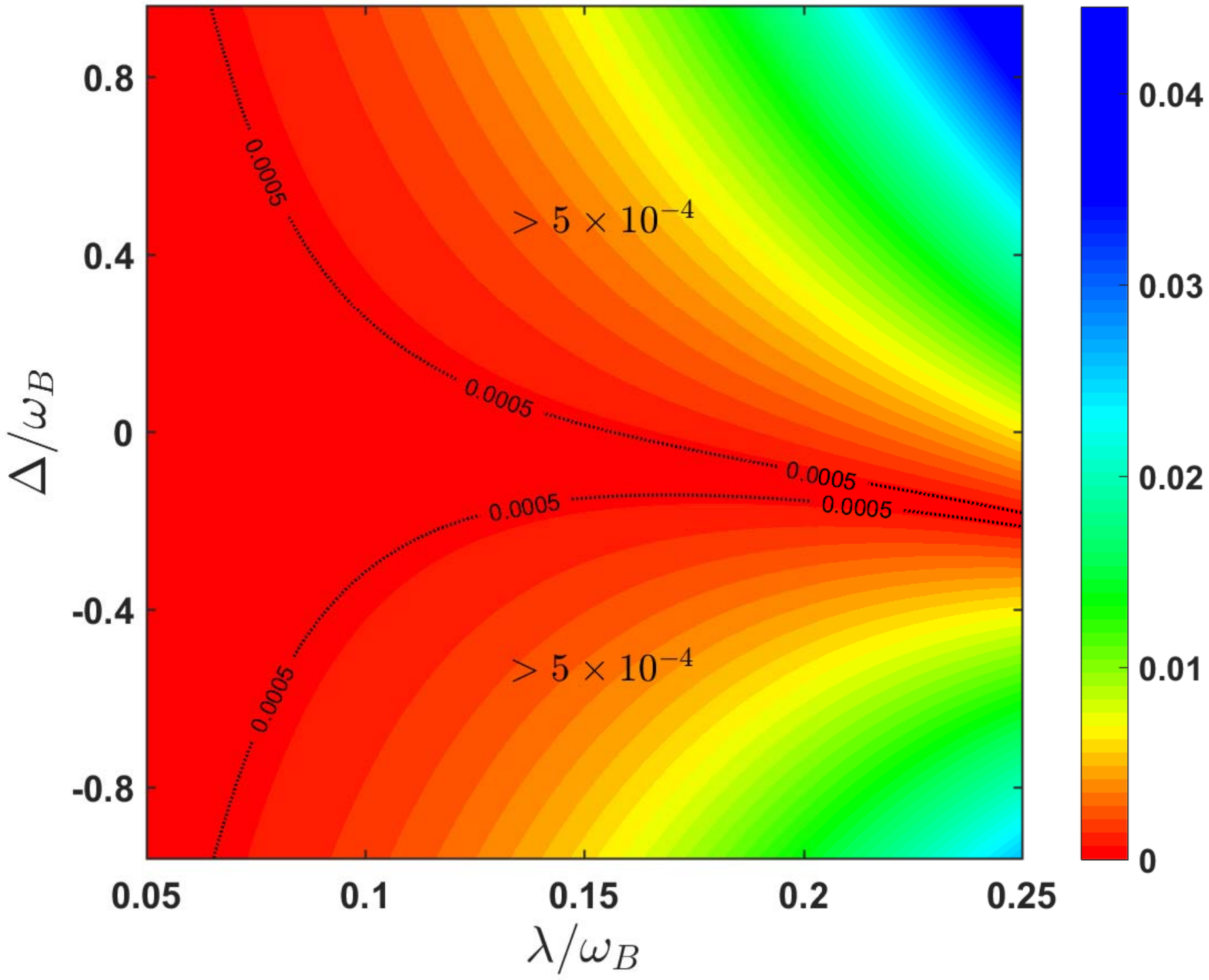}\\
  \caption{The effective coupling rate between the states $|g,g,1\rangle$ and $|e,e,0\rangle$ versus $\lambda/\omega_B$ and $\Delta/\omega_B$. The dashed lines correspond to the parameter regime for implementing the energy splitting $2\Omega_{\rm eff}/\omega_B=5\times 10^{-4}$. The system parameters used here are: $\omega_A=\omega_B$, $\omega_{fg}=(\omega_B-\Delta)/2$.}\label{fig7}
\end{figure}

In Fig.\,\ref{fig6}, we present the dependence of the maximum difference between $P^e_{AB}$ and $\langle a^{\dagger}a\rangle$ in a period of time on the coupling strength $\lambda/\omega_B$, which indicates the optimal regime (i.e., the green circles in Fig.\,\ref{fig6}) for obtaining single-photon-induced two qubit excitation in our model. We find that, both the coupling strength  $\lambda/\omega_B$ and the anharmonicity of qubit A have an influence on the maximum difference. Moreover, with the increase of qubit-cavity coupling strength, the maximum difference shows the trend of increasing first and then decreasing. We also find that the maximum difference is close to zero when $\lambda/\omega_B$ is very small. The reason is that, the excitation exchange has not yet started, but the system has been dissipated due to high dissipation rate and low value of the energy splitting $2\Omega_{\rm eff}/\omega_B$ between the states $|g,g,1\rangle$ and $|e,e,0\rangle$. We show the dependence of $2\Omega_{\rm eff}/\omega_B$ on $\lambda/\omega_B$ and $\Delta/\omega_B$ in Fig.7. The dashed lines indicate the effective coupling rate with a value $2\Omega_{\rm eff}/\omega_B=5\times 10^{-4}$. We see that, in the regime with a small coupling strength $\lambda/\omega_B$, the effective coupling rate has weak values, which correspond to small values of the maximum difference in Fig. 6. It is also seen that the maximum value of $2\Omega_{\rm eff}/\omega_B$ is obtained at the region of large coupling strength $\lambda/\omega_B$ and anharmonicity $\Delta/\omega_B$ (blue area). However, when the coupling strength is too high, the eigenstates of the Hamiltonian $H$ will be heavily dressed so that the resonance coupling occurred between the states $|g,g,1\rangle$ and $|e,e,0\rangle$ is difficult~\cite{ref17}. So the curves of the maximum difference show a downward trend with the increase of coupling strength in Fig.6. For some values of $\Delta/\omega_B$, the effective coupling rate is very small even when the system is in the large coupling strength region. This is because the contributions of higher order transition processes are main here. Then, the optimal regime shown in Fig.\,\ref{fig6} originally comes from the competition between the transition $|g,g,1\rangle\rightarrow|e,e,0\rangle$ decided by the coupling strength $\lambda/\omega_B$ and the dissipation of the system. From Fig.\,\ref{fig6}, we also observe that this optimal regime is controllable by manipulating the anhamonicity of the qubit A, which is experimentally accessible~\cite{ref59}.

\section{Conclusion}
In summary, we have proposed a method for achieving single-photon-induced two qubits excitation in a cavity QED system whose Hamiltonian has conserved $Z^2$ symmetry. The considered system includes a resonator and two qubits, and one of qubits has three levels due to its anharmonicity. We have studied the influence of system parameters including this anharmonicity on the realization of single-photon-induced two-qubit excitation, and the optimal regime is identified. We have also shown that the influence of the rate of cavity field damping and qubits decay on the single-photon-induced two-qubit excitation by the master equation approach. The competition between the state transition and system dissipation ultimately present the optimal regime for obtaining single-photon-induced two-qubit excitation in practice. This work provides a new perspective for the interaction between light and matter and has potential applications in quantum information science.
\begin{acknowledgments}
This work is supported by the National Key Research and Development Program of China grant
2016YFA0301203, the National Science Foundation of China (Grant Nos. 11374116, 11574104 and 11375067).
\end{acknowledgments}

%
%
%
%
%

\end{document}